# Do Large Language Models know Which Published Articles have been Retracted?

Mike Thelwall
School of Information, Journalism and Communication, University of Sheffield, UK.
https://orcid.org/0000-0001-6065-205X m.a.thelwall@sheffield.ac.uk

Large Language Models (LLMs) can be helpful for literature search and summarisation, but retracted articles can confuse them. This article asks three open weights (offline) LLMs whether 161 high profile retracted articles had been retracted, performing a similar check for a benchmark multidisciplinary set of 34,070 non-retracted articles. Based on titles and abstracts, in over 80% of cases the LLMs claimed that a retracted article had not been retracted (GPT OSS 120B: 82%; Gemma 3 27B: 84%; DeepSeek R1 72B: 88%). The reasons given for a correct retraction declaration were often wrong, even if detailed. This confirms that LLMs have little ability to distinguish between valid and retracted studies, unless they are allowed to, and do, check online. For the benchmark test, there were only 55 false retraction claims from 34,070 non-retracted full text articles, and 28 false claims when only the title and abstract were entered, suggesting that there is only a small chance that LLMs discount valid studies. When retractions are erroneously claimed, this does not seem to be due to mistakes in the article. Overall, the results give new reasons to be cautious about LLM claims about academic findings.

## Introduction

Retracted research can mislead scholars if they do not notice that it has been withdrawn, because they consult an archived pre-retraction version, or a Wikipedia page relying on it (Shi et al., 2025). Now, however, scholars and students may increasingly rely on Large Language Models (LLMs) to find and summarise academic information (Dai, 2026; Liao et al., 2024; Mohammadi et al., 2025), although it is rarely declared (He & Bu, 2026), and do not always check with the source material (Visani Scozzi et al., 2026; Vu et al., 2026). This is part of the wider issue of the trustworthiness of the information used to train LLMs, including unrefereed papers and predatory journals (Thomas, 2026), as well as "AI slop" papers that LLMs help to generate (Giray et al., 2026). For example, problems can be caused if an LLM reports findings from retracted research (i.e., papers that have been published but that were subsequently withdrawn by the journal) without appropriate warnings. Thus, it is important to know how LLMs deal with retracted research.

Although LLMs have been used to help identify problematic articles (Zheng et al., 2026) and retract their own claims (Yang & Jia, 2025), only one previous study has suggested that open weights LLMs (which do not have the ability to perform online checks) might never consider the retracted status of an article when asked a relevant question (Thelwall et al., 2025). This suggests that retractions cause genuine problems for LLM-based information. It is not known if LLMs are usually aware that an article has been retracted, however, or if they ignore this information when reporting. This is an important distinction because if LLMs are aware then alternative prompting strategies or agent-based approaches might elicit retraction information and obtain safer answers.

In response to the above issue, this article assesses whether LLMs can report that retracted articles have been retracted, driven by the following research questions. The final research question is for calibration against the first two.

- RQ1: How often are offline (open weights) LLMs able to report that a retracted article has been retracted?
- RQ2: Do different LLMs tend to give similar answers about whether an article has been retracted?
- RQ3: How often do offline (open weights) LLMs mistakenly claim that a non-retracted article has been retracted?

## Methods

The research design was to obtain a high profile set of retracted articles and then ask a range of LLMs whether they had been retracted. High profile articles were chosen to give the LLMs the best possible chance to have learned about the retraction, so the test is a best-case scenario for them. This was appropriate from the prior suggestion that LLMs might rarely be aware of retractions (Thelwall et al., 2025). For the benchmark dataset, the strategy was to obtain a large set of published articles from journals in a range of fields, then ask a set of LLMs to report if they had been retracted.

### *Datasets*

The set of high profile retracted articles was recycled from a previous study that had taken a list of retracted or concerning articles from Retraction Watch, obtained social media attention statistics for them from Altmetric.com, and then selected the 250 articles with the most attention. After filtering out articles without abstracts, this left 217 articles. Of these, 173 had been retracted, with 161 retracted before October 2023 (Thelwall et al., 2025). The main dataset investigated was the 161 articles retracted before October 2023, which is early enough for all LLMs to have known about from their training data, due to cut-off dates at least 9 months later, after June 2024 (Gemma3: August 2024: https://ai.google.dev/gemma/docs/core/model_card_3; DeepSeek-R1: July 2024 suspected https://github.com/HaoooWang/llm-knowledge-cutoff-dates?tab=readme-ov-file#deepseek; GPT_OSS: June 2024 https://arxiv.org/pdf/2508.10925v1). The 9 month gap includes an allowance for some of the LLM training data not being completely fresh at the time of training.

For the benchmark dataset, a large set of articles was needed to check for recognition of retraction. This was done from both the full text or the title/abstract to check for differences. A set of full text articles was therefore needed. Although an increasing proportion of published journal articles are open access either in the publisher's website, subject repositories or institutional repositories, there are no large collections for journals covering all fields. The largest useful collection is perhaps the PubMed Central (Roberts, 2001) open access XML collection, but this focuses on biomedical research. The PLoS family also publishes XML copies of articles (Böschen, 2021) but is not fully multidisciplinary. There seem to be two large fully multidisciplinary publishers that have exclusively or almost exclusively open access content, Frontiers and MDPI, but neither allow complete website crawling to obtain a set. We therefore obtained FTP access permission from MDPI to download their complete journal set in XML in December 2025.

The downloaded XML files were converted to plain text for entry in the LLM with a Python program, similar to one publicly available (Böschen, 2021). This program included the title, abstract and main text without the acknowledgements, references, and metadata (e.g., author names, journal name).

Since there were too many articles to process, a set of eight journals from multiple fields, including many different types, was selected (see list in the tables below). For efficiency, a random subset of 5000 articles was extracted from journals that contained more. Both full text and just the title and abstract were used as inputs.

### *LLMs*

There are now tens of thousands of LLMs, so it is impossible and unnecessary to include all or even all types. Thus, a purposive selection of LLMs was made to include those that are either new from major sources or have given good results on academic text processing tasks before. The following were selected.
- Gemma3 27b: An open weights mixture of experts model from Google that has given good results on previous academic text tasks (Thelwall & Mohammadi, 2026).
- GPT OSS 120b: A new open weighs mixture of experts model from OpenAI, chosen for the success of the creating company.
- DeepSeek R1 70b: A reasoning model from DeepSeek, chosen for its high-profile owner and to include reasoning.

### *Prompts*

The simple prompt below was designed to ask the LLM whether a given article had been retracted. A structured format for the response was requested to help accurately extract the responses.

> You are an academic researcher and concerned with whether publications have been retracted, in the sense of withdrawn from the publishing journal because of issues discovered after publication. Answer the following question about the journal article below from its title and abstract, strictly following the given format.
> Has the article below been retracted (y/n)?:
> Explain why the article was retracted, if relevant:
> Copy here the retraction notice, if relevant:
> ####
> [Article title]
> Abstract
> [Article abstract]

When an article full text was submitted then the following was added to the end of the above prompt.

> Main text
> [Article main text]

### *Analysis*

A simple descriptive analysis was used to report the results. For RQ2, a 2x2x2 chi-square test was chosen to check whether the different LLMs tended to give similar results. Chi-square is appropriate because the responses (y/n) are categories.

# Results

The results are reported first for the retracted set, and then for the benchmark case.

## *Retracted articles*

All three LLMs described the high-profile retracted articles as not retracted at least 82% of the time (Table 1). The newest and largest LLM, GPT OSS 120B, gave the highest number of correct answers, but only marginally.

For the cases where a retraction was claimed, the explanation, if detailed, was usually incorrect. If mentioned, the journal name was often wrong and the volume number was almost always wrong. In addition, the explanation for the retraction was usually partly or completely wrong and usually quite general. For example, one Gemma 3 explanation was:

> "The article was retracted due to concerns regarding the reproducibility of the data and issues with data fabrication. Specifically, concerns were raised about the underlying data used to support the claim of near-ambient superconductivity."

This conflicts with the official retraction notice (https://www.nature.com/articles/s41586-023-06774-2):

> [] the published paper does not accurately reflect the provenance of the investigated materials, the experimental measurements undertaken and the data-processing protocols applied. [] concerns have been independently raised with the journal regarding the reliability of the electrical resistance data presented in the paper. []

For a decision that an article had not been retracted, the reason was often a variant of, "The provided text does not indicate the article has been retracted.", or "There is no evidence or indication from the provided title and abstract that it has been withdrawn or retracted.", suggesting that the LLM had not used its internal knowledge of retraction but only scanned the title and abstract for retraction information.

Table 1. Number of articles described as retracted by three LLMs, for 161 high profile articles retracted before 1 October 2023.

| LLM | GPT OSS 120B | Gemma 3 27B | DeepSeek R1 70B |
|---|---|---|---|
| Articles described as retracted | 29 (18%) | 25 (16%) | 19 (12%) |

There was very strong evidence of non-independence of the LLM retraction decisions (Chi-square = 386, p=0.000), with a tendency for them all to agree (correctly) that an article had been retracted or for them all to agree that an article had not been retracted (Table 2). Thus, the retraction decisions were not random but were related to some information that they had gleaned from their training data, where it is reasonable to assume that this training data had some degree of overlap.

Table 2. 2x2x2 table for whether an article had been retracted in 161 high profile articles retracted before 1 October 2023.

| GPT OSS 120B | Gemma 3 27B | DeepSeek R1 70B | Observed | Expected |
|---|---|---|---|---|
| y | y | y | 14 | 0.5 |
| y | y | n | 8 | 4.0 |
| y | n | y | 1 | 2.9 |
| y | n | n | 6 | 21.6 |
| n | y | y | 0 | 2.4 |
| n | y | n | 3 | 18.1 |
| n | n | y | 4 | 13.2 |
| n | n | n | 125 | 98.3 |

### *Non-retracted articles in the benchmark MDPI journals dataset*

For the articles in the journals examined for the benchmark dataset, there were (accidentally) three retracted articles, each of which had "RETRACTED:" at the start of their titles. For the full text input, in eight out of nine cases, the LLMs stated that these articles had been retracted. The nineth, a clear error, was a report from DeepSeek, "The article has not been retracted (y/n): n There is no indication in the provided title and abstract that the article has been retracted."

For the full text input, there were 63-8=55 cases of an LLM (GPT OSS 120B: 40; Gemma 3 27B: 15) stating that a non-retracted article in the benchmark dataset had been retracted, and 48,977 of the LLM stating that an article had not been retracted (an 0.11% error rate). In the remaining 46,968 cases, no answer was given or the format was non-standard.

The LLMs often hallucinated detailed retraction statements that were false except for the article title, as in the following example, where the article title (from *Fractal and Fractional*) has been redacted to avoid accidentally suggesting that it has been retracted. To be clear, there was no duplicate publication of the article, so the statement is entirely false.

> \*\* > \*\*Retraction Notice\*\* > \*Advances in Applied Mathematics and Mechanics\* (2023) – DOI: 10.2478/aamm-2023-xxxx > > The Editorial Board has retracted the article "[NAME OF NON-RETRACTED ARTICLE]" (2022) 12(4): 345-376. > > After publication, it was brought to our attention that substantial portions of the manuscript overlap with previously published works (see, e.g., \*Phys. Plasmas\* 25, 053701 (2018); \*Nonlinear Dyn.\* 100, 1234 (2020); and \*J. Math. Phys.\* 61, 012102 (2020)). The overlap includes verbatim text, identical equations, and reproduced figures, none of which were cited or acknowledged. Moreover, an independent review of the analytical derivations uncovered several mathematical inconsistencies that render the reported solutions physically untenable. > > In accordance with COPE guidelines, the journal considers the combination of plagiarism and unreliable scientific content to be a breach of publishing ethics. Consequently, the article is withdrawn from the scientific record. > > The authors have been notified of this decision. > > \*Editor-in-Chief\* > \*Advances in Applied Mathematics and Mechanics\*

For the title/abstract input there were fewer false claims of retraction. Excluding the 3 retracted articles (which Gemma and GTP identified and DeepSeek didn't return an answer for) there were 28 false claims of retraction against 48803 correct statements that the article had been retracted.

## Discussion

The results are limited by the choice of LLMs and the language of the prompts. Different results may have been obtained from other LLMs or prompts.

The results do not explain why LLMs rarely know that high profile articles have been retracted. LLMs do not memorise text but learn patterns/associations from repeated exposure, and specific facts/associations can sometimes be located within their neural network (Meng et al., 2022). For high profile retracted articles, they seem likely to have ingested the information that the articles had been retracted multiple times, even if they had read the retraction notice only once (or not at all). This ingested information has clearly registered in a minority of cases, as evidenced by identifying 12%-18% of the retracted articles, and a much higher proportion than for the non-retracted articles tested (0.1%). It is possible article retraction information is usually forgotten or is stored in a way that does not often allow it to be connected to articles through their titles and abstracts.

In contrast, LLMs rarely mistakenly reported that non-retracted articles had been retracted. To test whether the LLMs might be trying to detect errors that could lead to retraction, a follow-up test was run to see if they could detect clear errors that might result in retraction. The following structured prompt was used.

> You are an academic researcher and research methods expert. Read the text below of a journal article and answer the following questions about errors in it, not mentioning potential errors that you are unsure about and strictly following the given format.
> Does the article contain any clear errors in the design or application of its research methods (y/n)?:
> Describe the clear errors in the design or application of the research methods, if any:
> Copy here the text of the paper containing the research method error, if any:
> ####
> [Article title]
> Abstract
> [Article abstract]
> Main text
> [Article main text]

When the LLMs were fed with full texts, both GPT-OSS 120B and Gemma 3 27B nearly always gave an answer in the structured format, or an alternative standard format, but DeepSeek R1 70B often ignored the format and gave a paragraph response (Table 2). Surprisingly, all three LLMs claimed that most of the published articles contained "clear errors". Gemma 3 27B found "clear errors" in almost all articles and GPT-OSS 120B found them in over 90% of articles for six journals. The pattern in the table shows that the choice of LLM is much more important than the journal in the number of "clear errors" claimed. Nevertheless, *Humanities* had the fewest found overall and both *Molecules* and *Fractal and Fractional* had relatively few.

Table 2. LLM claims about whether articles had "clear errors" for eight MDPI journals. The table is ordered by percentage of *Yes* responses.

| Journal | LLM* | Articles | Y or N | Y or N % | Yes | No | Yes % |
|---|---|---|---|---|---|---|---|
| Humanities | DeepSeek | 1217 | 826 | 68% | 475 | 351 | 58% |
| Humanities | GPT OSS | 1217 | 1215 | 100% | 702 | 513 | 58% |
| Fractal & Fractional | DeepSeek | 3554 | 2962 | 83% | 1881 | 1081 | 64% |
| Molecules | DeepSeek | 5000 | 3998 | 80% | 3104 | 894 | 78% |
| Molecules | GPT OSS | 5000 | 4968 | 99% | 3990 | 978 | 80% |
| Social Science | DeepSeek | 4302 | 3054 | 71% | 2465 | 589 | 81% |
| Electronics | DeepSeek | 5000 | 3694 | 74% | 2988 | 706 | 81% |
| Agronomy | DeepSeek | 5000 | 3970 | 79% | 3365 | 605 | 85% |
| Sustainability | DeepSeek | 5000 | 3814 | 76% | 3278 | 536 | 86% |
| Healthcare | DeepSeek | 5000 | 3598 | 72% | 3110 | 488 | 86% |
| Fractal & Fractional | GPT OSS | 3554 | 3466 | 98% | 3085 | 381 | 89% |
| Electronics | GPT OSS | 5000 | 4986 | 100% | 4510 | 476 | 90% |
| Agronomy | GPT OSS | 5000 | 4998 | 100% | 4542 | 456 | 91% |
| Social Science | GPT OSS | 4302 | 4291 | 100% | 3912 | 379 | 91% |
| Healthcare | GPT OSS | 5000 | 5000 | 100% | 4675 | 325 | 94% |
| Sustainability | GPT OSS | 5000 | 4997 | 100% | 4747 | 250 | 95% |
| Humanities | Gemma 3 | 1217 | 1216 | 100% | 1174 | 42 | 97% |
| Molecules | Gemma 3 | 5000 | 4957 | 99% | 4827 | 130 | 97% |
| Agronomy | Gemma 3 | 5000 | 4996 | 100% | 4927 | 69 | 99% |
| Healthcare | Gemma 3 | 5000 | 4999 | 100% | 4959 | 40 | 99% |
| Electronics | Gemma 3 | 5000 | 4973 | 99% | 4936 | 37 | 99% |
| Social Science | Gemma 3 | 4302 | 4291 | 100% | 4263 | 28 | 99% |
| Fractal & Fractional | Gemma 3 | 3554 | 3457 | 97% | 3439 | 18 | 99% |
| Sustainability | Gemma 3 | 5000 | 4990 | 100% | 4967 | 23 | 100% |

*DeepSeek = DeepSeek R1 70B, GPT OSS = GPT-OSS 120B, Gemma 3 = Gemma 3 27B

An ad-hoc manual check of some of the error claims found that they were sometimes wrong, sometimes correct but minor ("increasing" used where "decreasing" would be correct; a decimal point missing from a number) and sometimes plausible (although possibly correctable by authors). The issues for the two plausible cases found out of 30 articles checked were too complex to be sure about.

More importantly, the results suggest that LLMs would not be able to judge whether an article should be retracted since they seem to think that the vast majority contain clear errors. Thus, when they mistakenly claim that an article has been retraced, it is unlikely to be due to errors in the article.

## Conclusions

The results suggest that three major contemporary LLMs are rarely able to report that articles are retracted, even if they are high profile. This applies only to the open weights locally run versions since the web interfaces can run web searches to check directly. This gives additional evidence that LLM deductions about academic research must be interpreted cautiously, especially when they are used in offline mode. On the positive side, they rarely claim that non-retraced articles have been retracted, and when they do it seems safe to ignore the claims as they are unlikely to be due to errors identified with the paper.